# Augmenting the spin properties of shallow implanted NV-centers by CVD-overgrowth

T. Staudacher[1], F. Ziem[1], L. Häussler[1], R. Stöhr[1], S. Steinert[1], F. Reinhard[1], J. Scharpf[2], A. Denisenko[1,a)], and J. Wrachtrup[1]

[1] 3. Physikalische Institut and Research Center SCoPE, Universität Stuttgart, D-70174 Stuttgart

[2] Inst. Electron Devices and Circuits, Universität Ulm, D-89069 Ulm

**Abstract**

The controlled scaling of diamond defect center based quantum registers relies on the ability to position NVs with high spatial resolution. Using ion implantation, shallow (< 10 nm) NVs can be placed with accuracy below 20nm, but generally show reduced spin properties compared to bulk NVs. We demonstrate the augmentation of spin properties for shallow implanted NV centers using an overgrowth technique. An increase of the coherence times up to an order of magnitude ($T_2$ = 250μs) was achieved. Dynamic decoupling of defects spins achieves ms decoherence times. The study marks a further step towards achieving strong coupling among defects positioned with nm precision.

One of the key challenges in experimental quantum information science is the identification of isolated quantum mechanical systems which exhibit long coherence times and can be manipulated and coupled in a scalable fashion. Single defects in diamond and especially the negatively charged nitrogen vacancy (NV) center are a perfect platform for studying the quantum dynamics of spin systems. The NV consists of a substitutional nitrogen atom with an adjacent vacancy and an extra electron attached to its complex. It has a long-lived spin triplet in its electronic ground state with coherence times ranging up to 3 ms under ambient conditions [1]. The NV spin can be prepared and detected by optical means, and microwave excitation can be used to control its spin state [2, 3, 4]. Quantum registers based on the coupling of one NV center to an electron spin of another proximal NV center [5, 6] as well as with nuclear spins of neighboring $^{13}C$ atoms [7, 8, 9, 10, 11] have already been demonstrated experimentally.

Of special interest in the context of large scale spin arrays is the generation of multiple strongly coupled NV centers, since this allows the performance of advanced quantum protocols. Since the

[a)] Electronic mail: a.denisenko@physik.uni-stuttgart.de

coupling strength decreases with the inverse cubic distance, close neighboring NVs with long coherence times are necessary.

Recently nanometer-precision depth control of NV center creation near the diamond surface was achieved using an *in-situ* nitrogen delta-doping technique, showing coherence times of $T_2 > 100 \mu s$ for NV centers in 5 nm distance to the diamond surface [12]. However, the fabrication of scalable quantum devices in diamond [13] requires the ability to reliably create NV centers on demand with high spatial resolution in all three dimensions.

Having regard to these criteria, ion implantation is the most powerful technique for placing impurity atoms into the diamond matrix. Ion implantation is proven to be very versatile providing a wide range of tuning possibilities via ion energy (i.e. implantation depth) or ion density (resulting in single NVs up to large NV ensembles). Using implantation masks, one can generate strongly coupled pairs of deep, long-lived NV centers suitable for entanglement experiments [6, 14]. However, the yield for such strongly coupled NV pairs is relatively low due to persistent straggling in the diamond lattice.

Although the depth of the resulting centers can be tuned by choosing the implantation energy of the nitrogen, ion straggling and ion channeling set an intrinsic limitation of spatial accuracy. Ion straggling results in a broadening of the ion-implanted volume and is due to the multiple scattering processes experienced by the implanted ion throughout the diamond lattice. Straggling typically increases with higher implantation energy, ordinarily from a few nanometers at keV range up to several tens or hundreds of nanometers at MeV energies. With regard to quantum information processing applications, it has been shown that the coherence time of the electron spin associated to the NV center decreases in the proximity to the surface [15]. Hence, one faces a trade-off situation between the high placement accuracy of shallow implanted NVs with their lower spin properties and deep implanted NVs with reduced placement accuracy but better spin properties. This trade-off situation can be illustrated by comparing 2-D distributions of ion tracks in diamond as generated by SRIM simulation [16] for nitrogen implantation with 2.5, 5 and 10 keV energies used in this study (Fig. 1d).

In this Letter, we demonstrate an augmentation of the spin properties of shallow implanted NV centers by growing an additional diamond layer over the implanted diamond area. For this $^{15}$N ions were implanted into a commercially available *Element Six* electronic grade ($[N]_S^0 < 5$ ppb) and (100)–oriented single crystal diamond substrate. Six individual fields of implanted nitrogen were formed by photolithography masking and using the kinetic energies of 2.5 keV, 5 keV, and 10 keV and two doses of $10^{10}$ cm$^{-2}$ and $10^{13}$ cm$^{-2}$, respectively, thus allowing ensemble as well as single NV measurements. The implanted diamond sample was then annealed at 850°C temperature in high vacuum (pressure $< 10^{-6}$ mbar) for 10 hours. Afterwards, the annealed sample was boiled in a 1:1:1 mixture of sulfuric, nitric, and perchloric acids to remove any graphitic contamination at the surface stabilizing the charge state of the shallow NV centers [17].

In the next step, a nominally undoped diamond layer was grown on the entire surface of the substrate by chemical vapor deposition (CVD) in a microwave-assisted plasma reactor system. The CVD growth process was performed in a hydrogen/methane gas mixture with 200/0.5 SCCM flow rates, substrate temperature of 740 °C and gas pressure of 10 mbar. The thickness of the overgrown CVD layer was approximately 30 nm according to the growth calibration. Prior to CVD growth, the surface of the implanted diamond was subjected *in-situ* to H-plasma treatment at the same temperature and gas pressure, but with no methane flow present. This processing step has been used previously to smooth as-polished surface of single crystal diamond substrates leading to e.g. improved electronic properties of field effect transistors with H-induced conductive channels [18]. At the same time, exposure of diamond to microwave H-plasma can etch a surface damaged layer [19]. An amorphous carbon layer extending from 2 nm up to 4 nm induced by low-energy ion sputtering process could thus be completely abrogated by etching the diamond surface with such a plasma treatment for 20 min at approximately 700°C [19].

Figure 1 shows a schematic cross section of the overgrown surface accompanied by 2-D photoluminescence (PL) images of the substrate- growth layer interface within the nitrogen-implanted and non-implanted areas. The PL intensity from the substrate-to-CVD layer interface was about 3 orders of magnitude lower as compared to that of the nitrogen-implanted areas. In addition, even single NV centers in the lowest density region ($10^{10}$cm$^{-2}$) were clearly visible over the background fluorescence. The corresponding PL spectrum in Fig. 1c exhibits a featureless broad band in the 550-to-700 nm wavelength range. The PL spectrum from the implanted areas in Fig. 1c shows the presence of NV$^-$ defects (zero-phonon line (ZPL) of 638 nm) and NV$^0$ defects (575 nm ZPL). After the CVD re-growth the integrated intensity of the NV$^0$ PL signal is approx. 5 to 10% of the NV$^-$ PL signal at the given experimental conditions (532 nm laser, 538 nm long pass filter).

Before and after the diamond overgrowth step, the spin relaxation properties of the implanted NV centers were characterized with commonly used optically detected magnetic resonance techniques (ODMR). Specifically, we measured the longitudinal relaxation time $T_1$ in the densely implanted area ($10^{13}$ cm$^{-2}$) using a widefield setup. This device bins the response of approximately 3.6 million NVs on a 60 x 60 μm$^2$ field of view and therefore allows for a fast acquisition of the generally long (i.e. ms) relaxation times. The transverse coherence time $T_2$ was measured on single NVs using a home-built confocal microscope [5]. We employed the standard Hahn echo pulse sequence [2, 8] followed by a mono-exponential fit of the envelope decay to quantify $T_2$ which provides a general benchmark for the quality of the implanted NV centers.

Before overgrowth, $T_1$ times ranging from 1 ms to 1.8 ms were observed in the shallow dense NV ensembles ($10^{13}$ cm$^{-2}$) for all implantation energies. $T_1$ measurements monitor the return to thermal distribution among the spin states from polarization to m$_s$ = 0, thereby establishing an upper limit for the transverse dephasing time achievable by decoupling pulse schemes [20]. Typically we observe longitudinal magnetization decaying on two time scales (insert in Fig. 2a), the $T_1$ values quoted here are the longer component of a bi-exponential fit (Fig. 2a). We choose this value as a benchmark, since, firstly, the longer component has a roughly three times larger amplitude than the short component and, secondly, this timescale is the relevant figure of merit for sensing and decoupling schemes.

Upon overgrowth, we measured an increase of $T_1$ by a factor of 1.7 for the 2.5 keV and 10 keV implantation areas and 2.1 for the 5 keV implantation. This severe increase in relaxation times could have two reasons: On the one hand, surface effects are likely to be reduced under the overgrown CVD layer. On the other hand, residual lattice defects from the implantation and neighboring NV centers might be etched away during the initial H-plasma treatment in the CVD reactor. We note that $T_1$ does not scale in any meaningful manner with the implantation energy (Fig. 2a). We believe this effect to be an artifact arising from a high systematic error on the implantation dose [21] as it might easily arise e.g. from a defocused ion beam. Our interpretation of the relative increase, however, should be valid though, since the $T_1$ measurements were taken at the same location before and after overgrowth.

The $T_2$ times measured on single NVs in the low density region ($10^{10}$ cm$^{-2}$) were generally short (i.e. few tens of µs) and relatively dispersed after the implantation. This broad distribution of coherence times is presumably caused by residual crystal damage intrinsic to the impact of the accelerated ions [5, 22] or surface related effects [15]. After the growing process, we observe an increase in the $T_2$ times for all implantation depths. The effect is most pronounced for the statistical data of the 2.5 keV $10^{10}$ cm$^{-2}$ implant (overall 36 single NV centers) on which we observe on average a relative increase in the coherence time of more than an order of magnitude. The Hahn echo decay for a representative single 2.5 keV implanted NV before and for a different NV of same implantation energy after the overgrowth process is shown in Fig. 3. The periodic collapses and revivals are induced by the $^{13}$C environment around the NV [8, 23]. Fig. 2 summarizes the change in the $T_1$ and $T_2$ times for the implanted NV centers before and after the growth process.

Surprisingly, the shallowest NVs (2.5 keV) display the highest $T_2$ times after overgrowth. This might be explained by back-etching of residual implantation damage during the initial H-plasma. This etching step could remove the entire damage layer for a shallow implant while it would only remove a small, shallow fraction of the damage for a deep implant. In this case, the remaining decoherence is likely due to remaining defects such as the nearest neighbor di-vacancy R4 (or W6) [24], which can be removed by high temperature annealing [22]. An additional high temperature annealing process (T > 1200°C) might therefore reduce the dispersion after the overgrowth and potentially increase the coherence times further. However, the distribution might equally arise from the etching step being inhomogeneous. Indeed, we observed the density of NV centers to vary on a scale of ~10 µm after the overgrowth both in the widefield and confocal measurements.

We were able to efficiently increase the transverse coherence times by employing dynamical decoupling with a series of n π-pulses using the CPMG-n sequences [25] (see insert in Fig. 3).

This indicates that the coherence times are limited by a slow fluctuating magnetic noise. Most likely, this noise is not only due to the $^{13}$C spin bath since this would allow for an even longer $T_2$ in the range of 500µs [26]. A more likely cause are slowly fluctuating paramagnetic defects remaining from the implantation or grown into the CVD layer. In the latter case, $T_2$ could be further increased by optimization of the growth process [12, 27]. However, by using a CPMG-40 sequence, we were able to extend the $T_2$ time of an overgrown 2.5 keV NV center to more than 1 ms. With this coherence time, a two-qubit gate could be realized between to NV centers separated by nearly 30 nm [6]. Note that this is much larger than the placement accuracy achievable with a 2.5 keV implant (Fig 1d), so that our technique allow for the deterministic creation of strongly coupled NV centers.

In summary, we have shown that $T_2$ of NVs created by shallow nitrogen implantation can be extended by an order of magnitude and $T_1$ can be approximately doubled via a subsequent overgrowth process. This method brings forward a new way to create multiple strongly coupled NV centers due to the reduced straggling at low implantation energies and subsequent increase of coherence by the increased separation from the diamond surface. It thereby circumvents the current drawbacks of both shallow and deep implanted NVs in regard to the strong coupling condition. Hence, this method potentially enables the generation of large scale spin-arrays and scalable quantum registers for quantum computation schemes.

Furthermore, on the way to highly sensitive magnetic sensors with nanometer spatial resolution based on dense NV ensembles, thin layers of shallow NVs with long coherence times are needed [28]. Additionally, relaxation based magnetic sensing schemes [29, 30] focusing on the longitudinal relaxation benefit from long $T_1$ times [28]. Improvement of the combined implantation and overgrowth technique presented here, i.e. application of only a few nanometers of ultrapure diamond on top of low energy implantation could enable the creation of quasi δ-shaped NV distributions associated with low implantation energies while providing relaxation times common to deeper implantations.

The authors would like to acknowledge financial support by the EU via SQUTEC and Diamant, as well as the DFG via the SFB/TR21, the research groups 1493 "Diamond quantum materials" and 1482 as well as the Volkswagen Foundation. The authors wish to thank Roman Kolesov, Fedor Jelezko, Kangwei Xia, Nicolas Götz, and Ingmar Jakobi for stimulating discussions and experimental help. We further thank Jan Meijer for help with the SRIM [16] simulation software.

# References

[1] N. Zhao, J. Honert, B. Schmid, J. Isoya, M. Markham, D. Twitchen, F. Jelezko, R.-B. Liu, H. Fedder, and J. Wrachtrup, arXiv:1204.6513 [cond-mat.mes-hall]

[2] F. Jelezko and J. Wrachtrup, Phys. Status Solidi A 203, 3207-3225 (2006)

[3] F. Jelezko, T. Gaebel, I. Popa, M. Domham, A. Gruber, and J. Wrachtrup, Phys. Rev. Lett. 93, 130501 (2004)

[4] G. D. Fuchs, V. V. Dobrovitski, D. M. Toyli, F. J. Heremans, and D.D. Awschalom, Science 11, 1520-1522 (2009)

[5] P. Neumann, R. Kolesov, B. Naydenov, J. Beck, F. Rempp, M. Steiner, V. Jacques, G. Balasubramanian, M. L. Markham, D. J. Twitchen, S. Pezzagna, J. Meijer, J. T. Twamley, F. Jelezko, and J. Wrachtrup, Nature Phys. 6, 249-253 (2010)

[6] F. Dolde, I. Jakobi, B. Naydenov, N. Zhao, S. Pezzagna, C. Trautmann, J. Meijer, P. Neumann, F. Jelezko, and J. Wrachtrup, to be published (2012)

[7] P. Neumann, N. Mizuochi, F. Rempp, P. Hemmer, H. Watanabe, S. Yamasaki, V. Jacques, T. Gaebel, F. Jelezko, and J. Wrachtrup, Science 6, 1326-1329 (2008)

[8] L. Childress, M. V. Gurudev Dutt, J. M. Taylor, A. S. Zibrov, F. Jelezko, J. Wrachtrup, P. R. Hemmer, and M. D. Lukin, Science 13, 281-285 (2006)

[9] P. Neumann, J. Beck, M. Steiner, F. Rempp, H. Fedder, P. R. Hemmer, J. Wrachtrup, and F. Jelezko, Science 30, 542-544 (2010)

[10] L. Jiang, J. S. Hodges, J. R. Maze, P. Maurer, J. M. Taylor, D. G. Cory, P. R. Hemmer, R. L. Walsworth, A. Yacoby, A. S. Zibrov, and M. D. Lukin, Science 9, 267-272 (2009)

[11] L. Robledo, L. Childress, H. Bernien, B. Hensen, P. F. A. Alkemade, and R. Hanson, Nature 477, 574-578 (2011)

[12] K. Ohno, F. J. Heremans, L. C. Bassett, B. A. Myers, D. M. Toyli, A. C. Blezynski Jayich, C. J. Palmstrom, and D. D. Awschalom, arXiv:1207.2784v1 [cond-mat.mtrl-sci]

[13] D. M. Toyli, C. D. Weis, G. D. Fuchs, T. Schenkel, and D. D. Awschalom, Nano Lett. 10(8), 3168-3172 (2010)

[14] S. Pezzagna, D. Rogalla, H.-W. Becker, I. Jakobi, F. Dolde, B. Naydenov, J. Wrachtrup, F. Jelezko, C. Trautmann, and J. Meijer, Phys. Status Solidi A 208, 2017-2022 (2011)

[15] B. K Ofori-Okai, S. Pezzagna, K. Chang, R. Schirhagl, Y. Tao, B. A. Moores, K. Groot-Berning, J. Meijer, C. L. Degen, Phys. Rev. B 86, 081406 (2012)

[16]  J. Ziegler, SRIM 2008, online at www.srim.org

[17]  M. Hauf, B. Grotz, B. Naydenov, M. Dankerl, S. Pezzagna, J. Meijer, F. Jelezko, J. Wrachtrup, M. Stutzmann, F.Reinhard, and J. A. Garrido, Phys. Rev. B, 081304(R) (2011)

[18]  E. Kohn and A. Denisenko, "Doped Diamond Electron Devices" in "CVD Diamond for Electronic Devices and Sensors", by R. Sussmann (Ed.), J. Wiley& Sons Publ., 2009, 596 pp.

[19]  A. Denisenko, A. Romanyuk, C. Pietzka, J. Scharpf, E. Kohn, J. Appl. Phys. 108 (2010) 074901.

[20]  B. Naydenov, F. Dolde, L. T. Hall, C. Shin, H. Fedder, L. C. L. Hollenberg, F. Jelezko, and J. Wrachtrup, Phys. Rev. B 83, 081201(R) (2011)

[21]  A. Jarmola, V. Acosta, K. Jensen, S. Chemerisov, and D. Budker, Phys. Rev. Lett. 108, 197601 (2012)

[22]  B. Naydenov, F. Reinhard, A. Lammle, V. Richter, R. Kalish, U. F. S. D'Haenens-Johansson, M. Newton, F. Jelezko, and J. Wrachtrup, Appl. Phys. Lett. 97, 242511 (2010)

[23]  J. R. Maze, P. L. Stanwix, J. S. Hodges, S. Hong, J. M. Taylor, P. Cappellaro, L. Jiang, M. V. Gurudev Dutt, E. Togan, A. S. Zibrov, A. Yacoby, R. L. Walsworth, and M. D. Lukin, Nature 455, 644-647 (2008)

[24]  D. Twitchen, M. E. Newton, J. M. Baker, T. R. Anthony, and W. F. Banholzer, Phys. Rev. B 59, 12900 (1999)

[25]  G. de Lange, Z. H. Wang, D. Ristè, V. V. Dobrovitski, and R. Hanson, Science 330, 60-63 (2010)

[26]  J. R. Maze, J. M. Taylor, and M. D. Lukin, Phys. Rev. B 78, 094303 (2008)

[27]  G. Balasubramanian, P. Neumann, D. Twitchen, M. Markham, R. Kolseov, N. Mizuochi, J. Isoya, J. Achard, J. Beck, J. Tisler, V. Jacques, P. R. Hemmer, F. Jelezko, and J. Wrachtrup, Nature Mater. 8, 383-387 (2009)

[28]  S. Steinert, F. Ziem, L. Hall, A. Zappe, M. Schweikert, A. Aird, G. Balasubramanian, L. Hollenberg, and J. Wrachtrup, to be submitted (2012)

[29]  J. H. Cole, L.C.L. Hollenberg, Nanotechnology 20, 495401 (2009)

[30]  L. T. Hall, J. H. Cole, C. D. Hill, and L. C. L. Hollenberg, Phys. Rev. Lett. 103, 220802 (2009)

# Figures

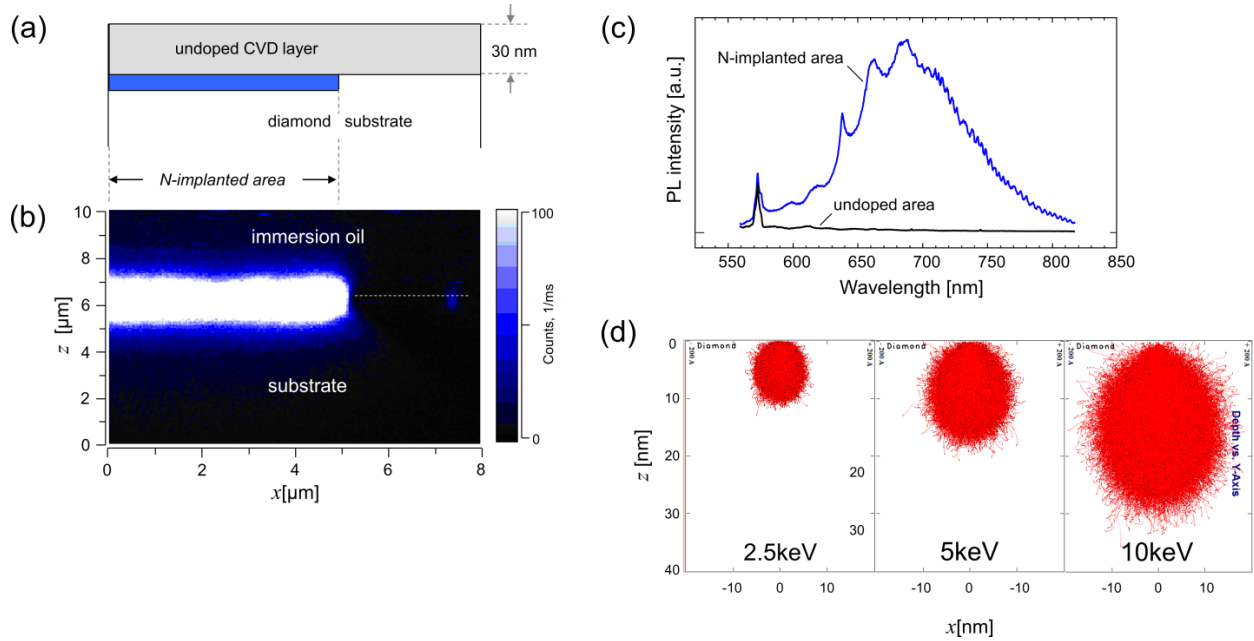

**FIG.1** A schematic cross section of the overgrown diamond surface (a) accompanied by a 2-D photoluminescence (PL) image (b) of nitrogen-implanted and non-implanted areas (vertical X-Z scan); (c) the corresponding PL spectra by 532 nm laser excitation with 538 nm long pass filter. (d) Estimate of the straggle during ion implantation using SRIM [16].

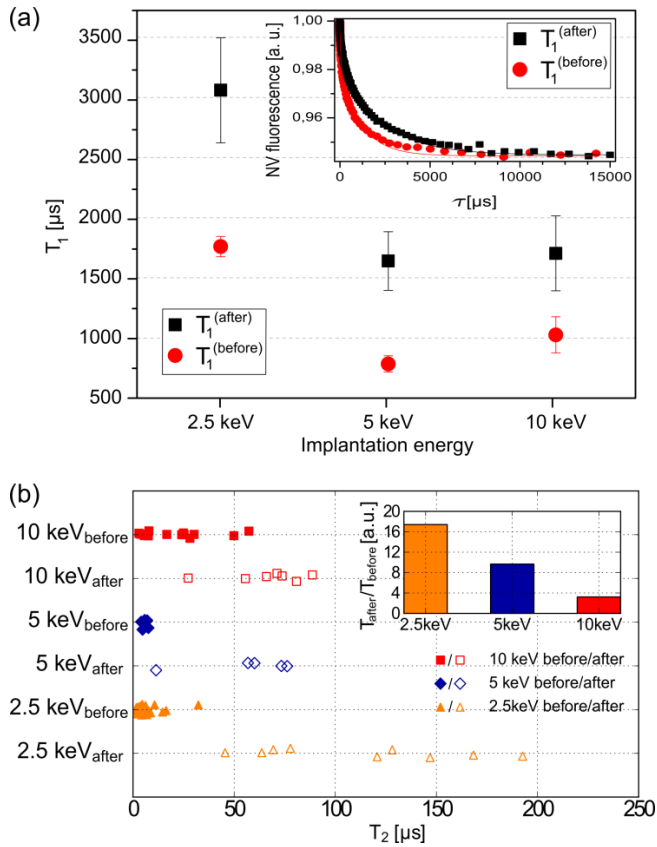

**FIG.2** Enhancement of the spin properties due to the overgrowth process. (a) Increase in the relaxation time $T_1$ for the ensemble, the error bars represent $1\sigma$ uncertainty. The inset shows the $T_1$ decay for the 10 keV $10^{13}$ cm$^{-2}$ ensemble before and after the overgrowth. The data was fitted with a bi-exponential decay. (b) Increase in the Hahn echo coherence time $T_2$ for single NV centers, each point is displaced by a small random offset along the vertical axis for better visibility. The inset shows the relative increase of the mean coherence times exceeding one order of magnitude for the 2.5 keV implanted NV centers.

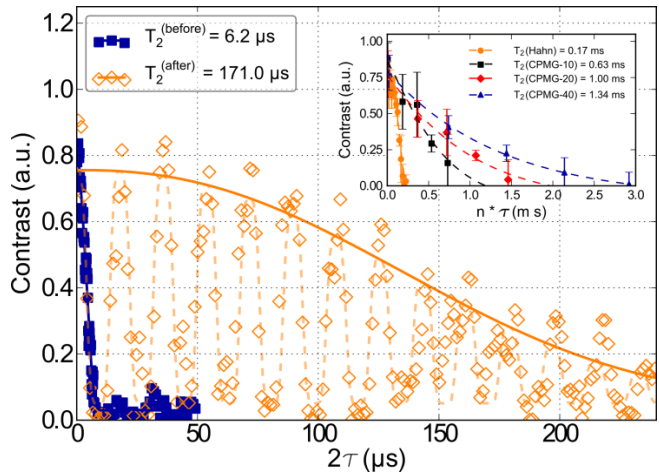

**FIG.3** Hahn echo decay for a representative single 2.5 keV implanted first NV before and for a second NV of same implantation energy after the overgrowth. Inset: Dynamical decoupling by a CPMG-n sequence prolongs the spin coherence of the overgrown second NV, exceeding 1 ms.